\documentclass{article}
\usepackage{float}
\usepackage[hidelinks]{hyperref} 

 \usepackage{graphicx} 

\usepackage[nonatbib,main,final]{neurips_2025}

\usepackage[backend=biber,style=ieee,maxnames=99]{biblatex}
\usepackage[utf8]{inputenc} 
\usepackage[T1]{fontenc}    
\usepackage{hyperref}       
\usepackage{url}            
\usepackage{booktabs}       
\usepackage{amsfonts}       
\usepackage{nicefrac}       
\usepackage{microtype}      
\usepackage{xcolor}         
\usepackage{booktabs}      
\usepackage{multirow}      
\usepackage{siunitx}       
\usepackage{makecell}      
\title{SoulX-Podcast: Towards Realistic Long-form Podcasts with Dialectal and Paralinguistic Diversity}

\author{%
\textbf{Hanke Xie}\textsuperscript{1,2}\thanks{Work done during an internship at Soul AI Lab. \texttt{hkxie@mail.nwpu.edu.cn}} \thanks{Equal contribution.} \and
\textbf{Haopeng Lin}\textsuperscript{2}\footnotemark[2] \and
\textbf{Wenxiao Cao}\textsuperscript{2} \and
\textbf{Dake Guo}\textsuperscript{1} \and
\textbf{Wenjie Tian}\textsuperscript{1} \and
\textbf{Jun Wu}\textsuperscript{2} \and
\textbf{Hanlin Wen}\textsuperscript{2} \and
\textbf{Ruixuan Shang}\textsuperscript{2} \and
\textbf{Hongmei Liu}\textsuperscript{2} \and
\textbf{Zhiqi Jiang}\textsuperscript{2} \and
\textbf{Yuepeng Jiang}\textsuperscript{1} \and
\textbf{Wenxi Chen}\textsuperscript{2,3}\footnotemark[1] \and
\textbf{Ruiqi Yan}\textsuperscript{2,3}\footnotemark[1] \and
\textbf{Jiale Qian}\textsuperscript{2} \and
\textbf{Yichao Yan}\textsuperscript{2} \and
\textbf{Shunshun Yin}\textsuperscript{2} \and
\textbf{Ming Tao}\textsuperscript{2} \and
\textbf{Xie Chen}\textsuperscript{3} \and
\textbf{Lei Xie}\textsuperscript{1}\thanks{Corresponding authors. \texttt{lxie@nwpu.edu.cn}, \texttt{wangxinsheng@soulapp.cn}} \and
\textbf{Xinsheng Wang}\textsuperscript{2}\footnotemark[3] \and
\textsuperscript{1}Audio, Speech and Language Processing Group (ASLP@NPU),\\
Northwestern Polytechnical University, Xi’an, China \\
\textsuperscript{2}Soul AI Lab, China \\
\textsuperscript{3}X-LANCE Lab, Shanghai Jiao Tong University, China
}

\begin{document}

\maketitle
\vspace{-2.5em}

\begin{abstract}

\vspace{-0.6em}

Recent advances in text-to-speech (TTS) synthesis have significantly improved speech expressiveness and naturalness. However, most existing systems are tailored for single-speaker synthesis and fall short in generating coherent multi-speaker conversational speech. This technical report presents SoulX-Podcast, a system designed for podcast-style multi-turn, multi-speaker dialogic speech generation, while also achieving state-of-the-art performance in conventional text-to-speech (TTS) tasks.
To meet the higher naturalness demands of multi-turn spoken dialogue, SoulX-Podcast integrates a range of paralinguistic controls and supports both Mandarin and English, as well as several Chinese dialects, including Sichuanese, Henanese, and Cantonese, enabling more personalized podcast-style speech generation. Experimental results demonstrate that SoulX-Podcast can continuously produce over 90 minutes of conversation with stable speaker timbre and smooth speaker transitions. Moreover, speakers exhibit contextually adaptive prosody, reflecting natural rhythm and intonation changes as dialogues progress. Across multiple evaluation metrics, SoulX-Podcast achieves state-of-the-art performance in both monologue TTS and multi-turn conversational speech synthesis.

\end{abstract}
\vspace{-1.1em}

\hspace*{0.09\textwidth}
\begin{minipage}{0.9\textwidth}
\small
\textbf{Demo page:} \href{https://soul-ailab.github.io/soulx-podcast/}{\textcolor{blue}{https://soul-ailab.github.io/soulx-podcast}} \\
\textbf{Source code:} \href{https://github.com/Soul-AILab/SoulX-Podcast}{\textcolor{blue}{https://github.com/Soul-AILab/SoulX-Podcast}}
\end{minipage}

\begin{figure}[hb]
    \centering
    \includegraphics[width=0.8\linewidth]{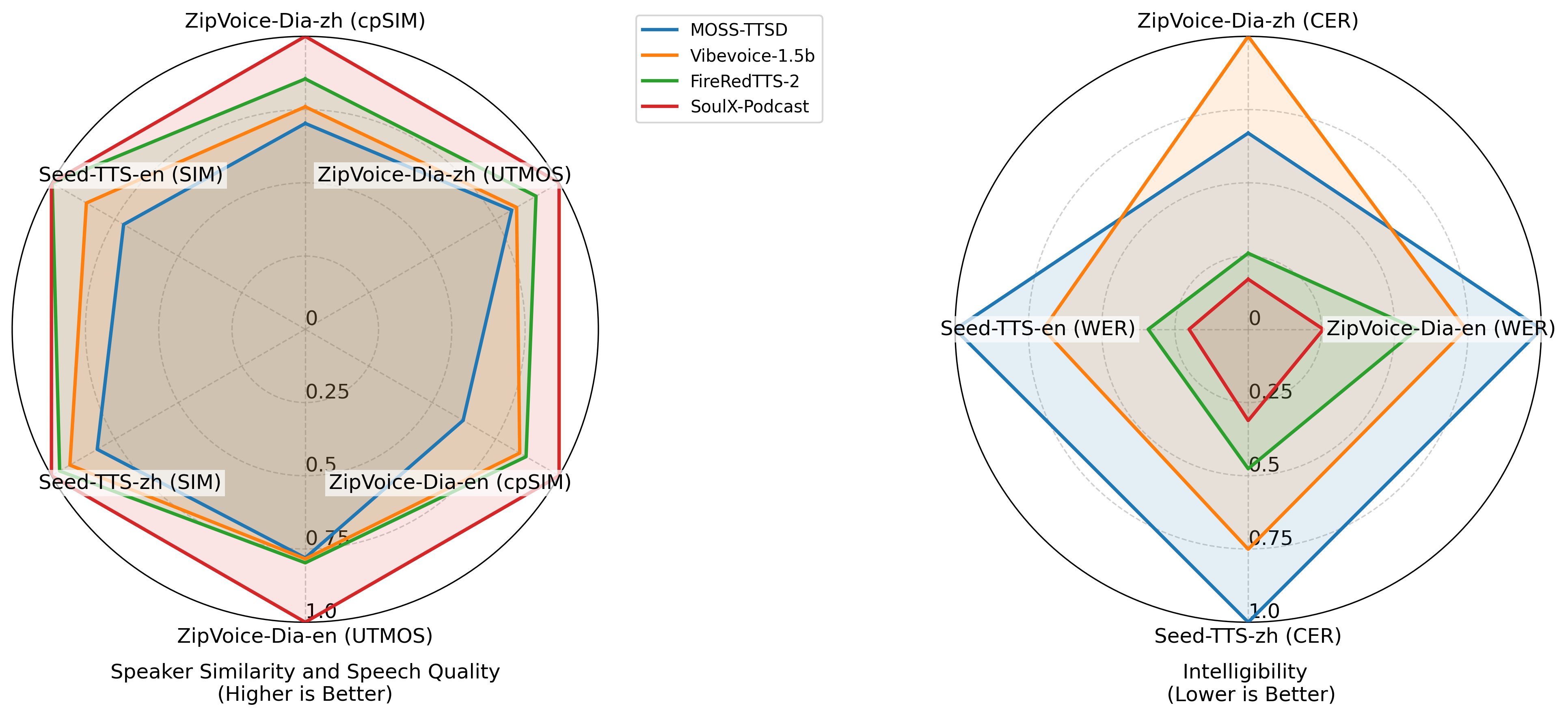}
    \caption{Performance of SoulX-Podcast.}
    \label{fig:soulxpodcast}
\end{figure}

\section{Introduction}

Building upon the generative power of large language models (LLMs), modern text-to-speech (TTS) systems have reached a point where they can produce speech that is nearly indistinguishable from human voices, achieving remarkable naturalness and zero-shot voice cloning performance~\cite{guo2024fireredtts, ye2025llasa, wang2025spark, deng2025indextts, du2025cosyvoice3}. However, most previous work has primarily focused on single-speaker speech generation. These systems, while effective in isolated speech tasks, struggle to maintain fluency and naturalness in multi-speaker, multi-turn conversation scenarios. In response to this gap, this technical report introduces \textbf{SoulX-Podcast}, a speech synthesis model specifically designed for seamless multi-speaker, multi-turn dialogues. To further enhance conversational realism and diversity, SoulX-Podcast incorporates robust support for various paralinguistic features and dialects, ensuring a more dynamic and natural dialogue experience.

Speech tokenization methods~\cite{huang2023repcodec, tao2024toneunit, xin2024bigcodec, ji2024wavtokenizer, wu2024ts3, ren2024fewer, li2024single, zheng2024freecodec, chen2025sac}, based on Vector Quantization (VQ)~\cite{van2017neural} or Finite Scalar Quantization~\cite{mentzer2023finite}, bridge the gap between continuous speech signals and discrete token-based large language models (LLMs). Vall-E~\cite{wang2023neural} is a pioneering speech generation system that leverages LLMs with discrete tokens and employs a tokenizer based
on residual vector quantization (RVQ)~\cite{encodec}. Specifically, tokens from the first layer are predicted by an autoregressive (AR) LLM, while the remaining tokens are generated using a non-autoregressive (NAR) model. Subsequent work, depending on the choice of tokenizer, can be categorized into several main modeling approaches: predicting a single stream of semantic tokens with LLMs and then generating acoustic features via flow matching~\cite{betker2023better, casanova2024xtts, anastassiou2024seed, cosyvoice, cosyvoice2, du2025cosyvoice3}; directly predicting acoustic tokens spanning multiple codebooks according to specific patterns~\cite{wang2024maskgct, xie2025fireredtts}; or directly predicting a single stream of acoustic tokens~\cite{ye2025llasa, wang2025spark}.

Most of the aforementioned research primarily focuses on monologue-style speech generation, overlooking the challenges of multi-speaker, multi-turn conversational synthesis. In contrast, conversational speech generation places higher demands on natural prosodic and rhythmic variation to ensure smooth and coherent dialogue flow. Recently, several studies have begun to explore this direction. For example, Covomix~\cite{zhang2024covomix} adopts a parallel-channel modeling strategy that simultaneously predicts the speech of different speakers through separate channels, while MoonCast~\cite{ju2025mooncast} and MOSS-TTSD~\cite{moss2025ttsd} merge dialogue text with speaker labels to generate integrated multi-speaker conversations. The latest FireRedTTS-2~\cite{xie2025fireredtts} model, on the other hand, produces speech from multiple speakers in an alternating manner. Although these methods achieve improved dialogue continuity and prosodic variation compared with standard TTS systems, their limited control over paralinguistic features still constrains the expressiveness and realism of the generated conversations.

In this work, we introduce SoulX-Podcast, a large language model–driven framework for long-form, multi-speaker, and multi-dialect podcast speech synthesis. SoulX-Podcast is designed to generate stable, coherent, and expressive podcast-style dialogic speech by effectively modeling dialectal variation, paralinguistic cues, and context-dependent prosody. The framework represents interleaved text–speech sequences, where speaker-labeled text and corresponding speech tokens are chronologically aligned, thereby facilitating the generation of long-form conversational audio with consistent quality and speaker similarity. Experimental results demonstrate that SoulX-Podcast delivers superior performance in multi-turn dialogue synthesis and exhibits strong generalization to conventional TTS tasks, highlighting its versatility across diverse speech generation scenarios.

In summary, SoulX-Podcast offers several distinctive features:
\begin{itemize}
    \item It supports \textbf{long-form, natural dialogue speech generation} with a variety of \textbf{paralinguistic labels}, achieving high fluency and coherence across extended multi-turn dialogues.

    \item In addition to \textbf{Mandarin} and \textbf{English}, SoulX-Podcast provides robust support for several Chinese dialects, including \textbf{Sichuanese}, \textbf{Henanese}, and \textbf{Cantonese}, enabling more diverse and personalized voice generation. Importantly, all of these dialects support \textbf{Cross-dialectal, zero-shot voice cloning}, allowing a single audio prompt to generate speech in any of the supported dialects.

    \item SoulX-Podcast demonstrates superior performance not only in multi-turn conversational speech synthesis but also in conventional TTS tasks, such as \textbf{voice cloning}, highlighting its effectiveness and versatility across diverse speech synthesis scenarios.
\end{itemize}



\section{Method}

Dialogue text–speech paired data that contain speaker identity correspondence are a necessary prerequisite for building a dialogue speech synthesis system. This section first introduces the data processing methods used in this work, including the handling of dialogue data and the annotation of dialectal and paralinguistic information. Subsequently, we present the specific algorithm of SoulX-Podcast.

\subsection{Data Processing}

\begin{figure}[h]
    \centering
    \includegraphics[width=0.8\linewidth]{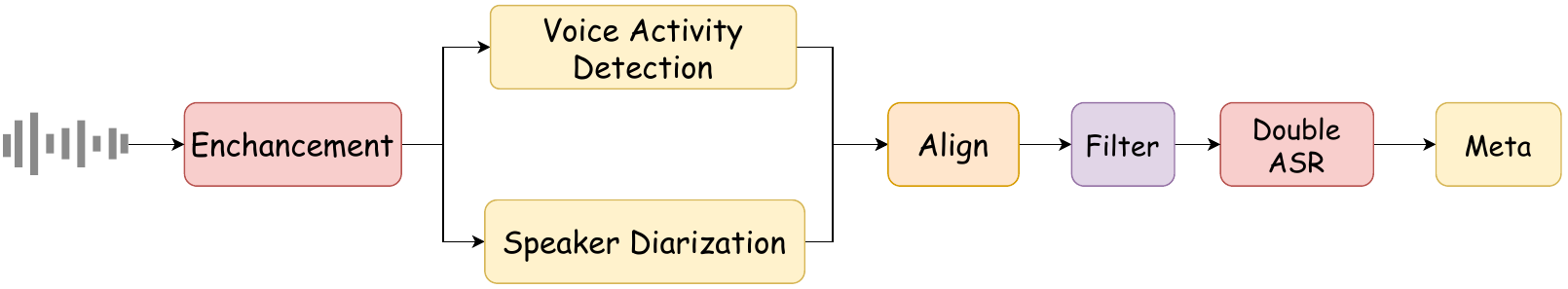}
    \caption{Processing pipeline for in-the-wild dialogue speech data.}
    \label{fig:datapipeline}
\end{figure}

In contrast to monologue speech, the processing of dialogue speech necessitates not only obtaining aligned transcripts but also distinguishing between speakers explicitly. As shown in Figure~\ref{fig:datapipeline}, the overall workflow comprises speech enhancement, audio segmentation and speaker diarization, text transcription, and quality filtering. Additionally, to facilitate paralinguistic and dialectal controllability, further information is extracted and annotated.

\subsubsection{Basic Processing Workflow}

\textbf{Audio Pre-processing.} 
In-the-wild dialogue recordings often contain background music or noise, which can adversely affect downstream tasks such as transcription or speaker diarization. To address this, we first apply a UVR-MDX-based\footnote{\url{https://github.com/seanghay/uvr-mdx-infer}} vocal separation tool to remove background audio and noise, and then normalize the resulting signals to a consistent amplitude.

\textbf{Segmentation and Diarization.} Processing long-form dialogue recordings (e.g., longer than 30 minutes) poses challenges for conventional speaker diarization. As speakers’ vocal characteristics and speaking states vary over time, diarization models may mistakenly assign multiple speaker identities to the same individual, leading to inconsistencies in speaker counts and boundary alignment.

To mitigate this issue, we first apply Voice Activity Detection (VAD)~\cite{SileroVAD} to segment long recordings into short utterances. These utterances are then concatenated into dialogue segments of approximately five minutes. During this process, we enforce a silence-duration constraint to prevent segment boundaries from crossing different sessions or long transitional silences: if the inter-utterance silence exceeds a predefined threshold, the adjacent utterances are treated as the end and start of separate segments.

Finally, we employ a \texttt{Sortformer}-based diarization model~\cite{park2024sortformer} to detect speaker boundaries and assign speaker labels, producing reliable speaker-turn annotations for subsequent processing.

\textbf{Quality Filtering.} 
Although the audio recordings were enhanced in the initial stage, some segments still exhibited suboptimal denoising results or inherently poor recording quality. To prevent such low-quality data from negatively impacting model training, we applied a series of filtering criteria to the dialogue segments, including signal-to-noise ratio (SNR) and perceptual quality estimated by DNSMOS~\cite{dnsmos}.

\textbf{Speech Recognition.}
Following the quality filtering process, we employed a dual-ASR transcription strategy to obtain reliable transcripts. Specifically, each utterance within a dialogue segment was transcribed by two independent ASR models. For Chinese speech, we used \texttt{Paraformer}\footnote{https://huggingface.co/funasr/Paraformer-large} and \texttt{Whisper}\footnote{https://huggingface.co/openai/whisper-large-v3}, while for English speech, we adopted \texttt{Parakeet}\footnote{https://huggingface.co/nvidia/parakeet-tdt-0.6b-v2} and \texttt{Whisper}.

For each utterance, two transcription results were obtained, and the Character Error Rate (CER) for Chinese or Word Error Rate (WER) for English was computed. Utterances with CER or WER below a predefined threshold were fully retained, with Paraformer outputs used as the final transcripts for Chinese and Whisper outputs used for English. For utterances whose CER or WER exceeded the threshold, only the textual transcripts were preserved, while the corresponding audio was discarded.

This strategy maintains dialogue completeness and textual consistency while minimizing the adverse impact of transcription errors on speech synthesis training, thereby achieving a better trade-off between data retention and transcription reliability.

\textbf{Speaker Purity Refinement.} To ensure speaker label consistency, we conducted a speaker-purity refinement based on speaker embedding clustering. For each dialogue segment, the embeddings of all utterances belonging to the same speaker were clustered, and utterances whose embeddings deviated excessively from the cluster centroid were identified as outliers. These outlier utterances were excluded from the audio data—only their transcriptions were retained. This strategy effectively mitigates potential speaker confusion during multi-turn dialogue synthesis while maximizing overall data retention. Here, we extract speaker embeddings with WavLM-large, finetuned on the speaker verification task~\cite{chen2022large}.
 
\vspace{-0.8em}

\subsubsection{Paralinguistic and Dialectal Data Annotation}

Paralinguistic cues, such as laughter and sighs, play a crucial role in enhancing the naturalness and expressiveness of dialogue. To enable controllable generation of such cues, we performed paralinguistic mining and annotation on the collected data.  Moreover, most previous speech synthesis research has primarily focused on Mandarin Chinese, while major Chinese dialects such as Cantonese and Sichuanese have received limited attention. To facilitate effective dialectal controllability, we further annotated the collected data with dialectal labels, enabling the model to capture and reproduce dialect-specific characteristics.

\textbf{Paralinguistic Annotation.} 
To ensure both large-scale coverage and fine-grained precision of paralinguistic labels, we design a two-stage refinement framework for data annotation. This framework combines high-throughput automated detection with model-assisted verification to achieve both efficiency and accuracy.

In the first stage, we employ language-specific ASR models fine-tuned for paralinguistic event detection to process the raw audio corpus. For Mandarin Chinese data, we use \texttt{Beats}~\cite{Chen2022beats} for coarse identification of nonverbal cues, while for English data, we adopt \texttt{Whisperd}~\cite{darefsky2024parakeet}. This stage efficiently filters out segments unlikely to contain relevant paralinguistic events.

In the second stage, the pre-annotated segments undergo model-driven verification and fine-grained labeling using the Gemini-2.5-Pro API~\cite{comanici2025gemini}. This multimodal model verifies the presence and category of paralinguistic events and generates precise, time-aligned annotations alongside the corresponding text.

Through this meticulous two-stage process, we obtain approximately 1,000 hours of high-quality speech with detailed paralinguistic annotations, providing a robust foundation for expressive and context-aware speech synthesis.

\textbf{Dialectal Annotation.} 
To efficiently collect dialectal speech data, we employed two complementary strategies. First, we collected publicly available recordings in specific Chinese dialects. Second, we trained a dialect identification model to retrieve and categorize dialectal utterances from the broader in-the-wild dataset.

For transcription, we observed that our standard pipeline performed suboptimally on dialectal speech. Accordingly, we leveraged the commercial Seed-ASR API~\footnote{https://docs.byteplus.com/zh-CN/docs/byteplusvoice/asrstreaming} to generate reliable transcripts. Using this approach, we obtained approximately 2,000 hours of Sichuanese, 1,000 hours of Cantonese, and 500 hours of Henanese speech.


\subsubsection{Corpus Overview}
Using the aforementioned methods, we ultimately obtained approximately \textbf{0.3 million hours} of high-quality, natural conversational speech. In addition, we curated roughly 1.0 million hours of monologue data, resulting in a total training dataset of approximately \textbf{1.3 million hours}.

\subsection{SoulX-Podcast}

\begin{figure}
    \centering
    \includegraphics[width=1\linewidth]{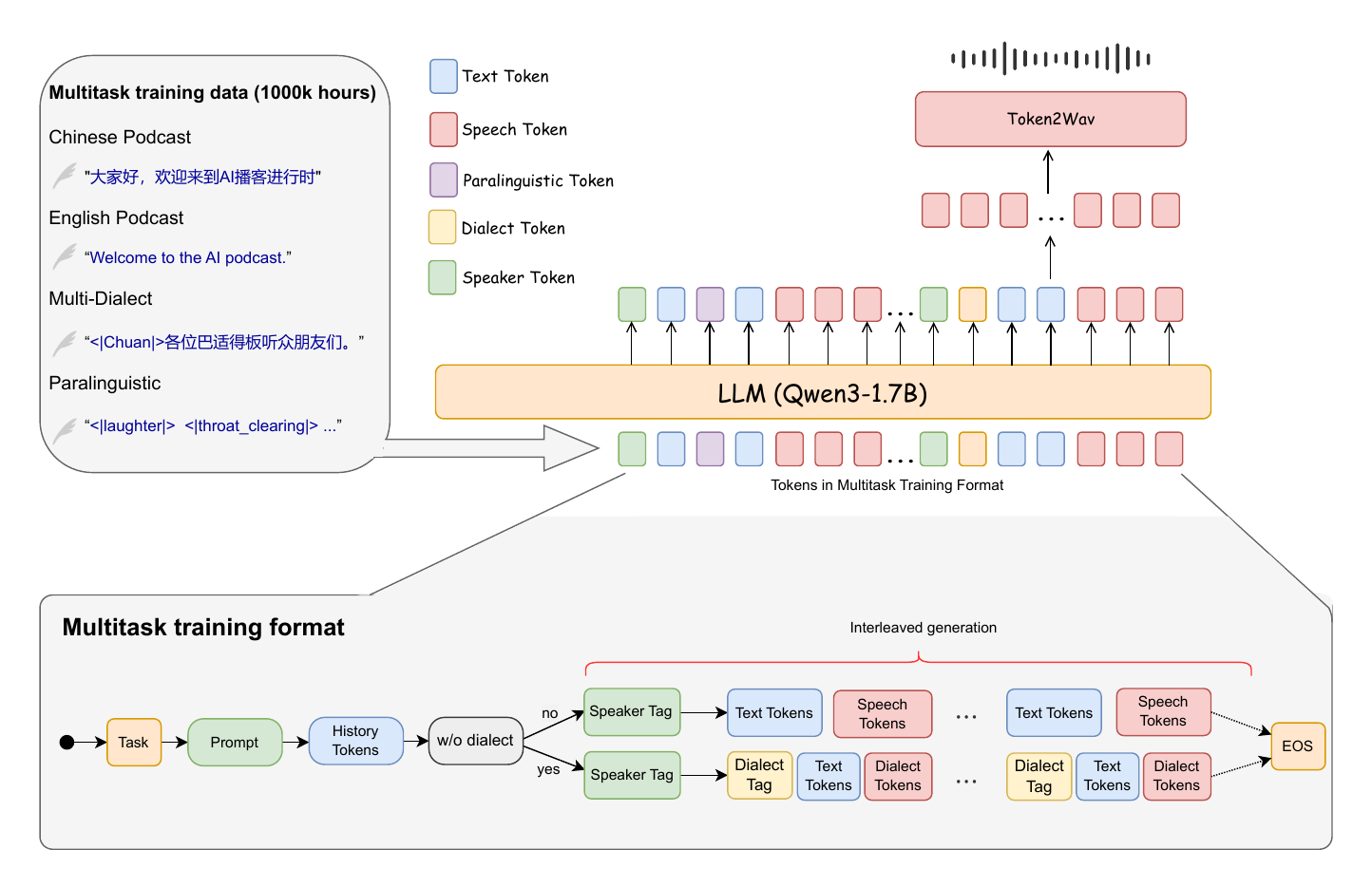}
    \caption{Overview of SoulX-Podcast}
    \label{fig:soulxpodcast}
\end{figure}

Following the CosyVoice series~\cite{cosyvoice2,du2025cosyvoice3}, SoulX-Podcast adopts a two-stage generative framework. Specifically, an LLM first predicts semantic tokens, which are then converted into acoustic features through flow matching and subsequently synthesized into waveform audio via a vocoder. The LLM backbone is the pre-trained Qwen3-1.7B model, whose text codebook is extended to accommodate both speech tokens and special tokens that encode paralinguistic and dialectal attributes.

\subsubsection{Token Organization}
To enable flexible, multi-turn dialogue generation, we adopt a text–speech interleaved sequence that allows sentence-by-sentence synthesis. Specifically, each speaker’s text tokens are followed by their corresponding speech tokens, which are then concatenated with the next speaker’s text and speech tokens in temporal order. Each utterance begins with a speaker token to indicate the speaker identity. Likewise, dialect control is achieved by inserting a dialect-specific token immediately after the speaker token, while paralinguistic cues (e.g., laughter, sighs) are treated as textual tokens and placed at their corresponding positions within the sequence.
An example with a dialect label is shown below: 

\texttt{<SPEAKER1><Sichuan><Text Tokens><Audio Tokens><SPEAKER2><Sichuan><Text Tokens><Audio Tokens><SPEAKER3><\ldots>}

\subsubsection{Training}
Dialogue speech data are relatively scarce compared to monologue speech. To effectively leverage heterogeneous data patterns and enhance performance in dialogue scenarios, we adopt a curriculum learning strategy.

In the first stage, the LLM backbone is initialized from Qwen3-1.7B~\footnote{https://huggingface.co/Qwen/Qwen3-1.7B} and trained on a mixture of monologue and dialogue data to acquire fundamental text-to-speech capabilities. Subsequently, the model is further trained on multi-speaker dialogue data in both Chinese and English, incorporating dialectal and paralinguistic elements. Since the amount of Chinese dialect data is significantly smaller than that of Mandarin and English, we perform additional fine-tuning on dialectal data to enhance the model’s dialectal capability, resulting in a podcast model specifically optimized for dialect generation.

To address the challenges of long-form audio generation, we introduce a context regularization mechanism that progressively drops historical speech tokens while retaining their textual context. This encourages the model to rely on semantic continuity rather than low-level acoustic memory, thereby improving coherence and stability in extended conversational synthesis.

\subsubsection{Inference}

During inference, we follow the token organization established in training: initial text and speech tokens from multiple speakers are interleaved, and the model autoregressively generates subsequent speech tokens in the same interleaved manner.  

\textbf{Cross-dialectal Voice Cloning.} For dialect generation, our goal is to enable cross-dialectal voice cloning. However, this is nontrivial. Unlike the clear orthographic differences between Chinese and English, various Chinese dialects—particularly Mandarin, Henanese, and Sichuanese—share an identical written form. Even Cantonese, though linguistically more distinct, still exhibits substantial textual overlap with Mandarin. Consequently, when the target text is highly similar to Mandarin and the speech prompt is also in Mandarin, the dialectal control signal becomes weak.

To address this issue and allow a Mandarin prompt to generate speech in any target dialect, we propose \textbf{Dialect-Guided Prompting (DGP)} inference strategy. Specifically, before generating a dialectal podcast, we prepend a short dialect-typical sentence—one that strongly reflects the target dialectal style—to the input text. This initial utterance effectively guides the model toward producing speech with the desired dialectal characteristics in subsequent generations.

\begin{figure}
    \centering
    \includegraphics[width=0.8\linewidth]{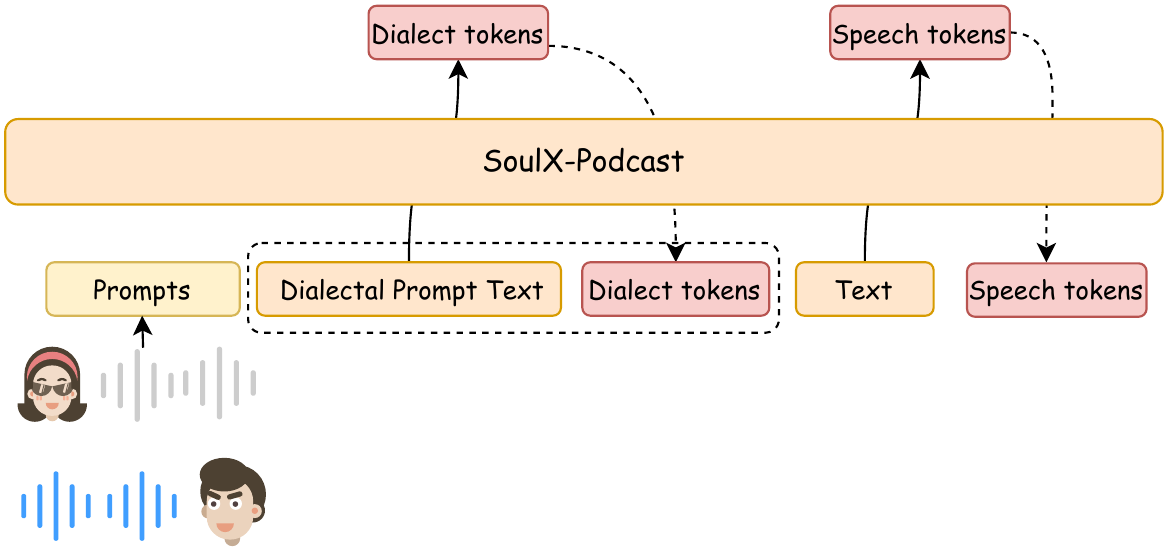}
    \caption{Inference procedure of SoulX-Podcast. The model supports Cross-dialectal prompting, where a Mandarin prompt can generate speech in target dialects with Dialect-Guided Prompting (DGP) method.}
    \label{fig:infer_of_soulxpodcast}
\end{figure}

\section{Performance of SoulX-Podcast}
\label{sc:results}

Although SoulX-Podcast is designed for multi-turn, multi-speaker dialogue synthesis, it is also capable of conventional monologue speech synthesis. Accordingly, we first compare its performance against SOTA TTS models on the standard monologue synthesis task. We then evaluate SoulX-Podcast’s capabilities in dialogue generation, as well as in paralinguistic and dialectal control.

\subsection{Monologue Speech Generation}

To evaluate the zero-shot (voice cloning) TTS capability of SoulX-Podcast, we assess its performance on Seed-TTS-eval and compare it with existing zero-shot TTS models. The results are summarized in Table~\ref{tab:seedtestset}, where speech intelligibility is measured using CER for Chinese and WER for English, and speaker similarity (SIM) is quantified via the cosine similarity of speaker embeddings, following the Seed-TTS-eval protocol\footnote{https://github.com/BytedanceSpeech/seed-tts-eval}. As can been seen, SoulX-Podcast demonstrates significant superiority in intelligibility for zero-shot monologue TTS scenarios. Specifically, SoulX-Podcast achieves lowest CER in the Chinese test set. In the English test set, SoulX-Podcast only seconds to F5-TTS. In terms of speaker similarity, SoulX-Podcast also achieves strong results. Specifically, on both the Chinese and English test sets, it ranks just behind Seed-TTS and MaskGCT, demonstrating its excellent performance in conventional zero-shot TTS.

\begin{table*}[!h]
    \centering
    \caption{
        TTS performance of different models on the Seed test sets (test-zh for Chinese, test-en for English). Arrows indicate the desired direction ($\downarrow$ = lower is better, $\uparrow$ = higher is better). Best values per column are in \textbf{bold}. {*} represents results reproduced from the official release.
    }
    \label{tab:seedtestset}
    \setlength{\tabcolsep}{6mm}
    \begin{tabular}{lcccc}
        \toprule
        \multirow{2}{*}{\textbf{Model}} &  \multicolumn{2}{c}{\textbf{test-zh}} & \multicolumn{2}{c}{\textbf{text-en}} \\
        \cmidrule(lr){2-3} \cmidrule(lr){4-5}
        & {\makecell{CER ($\downarrow$)}} & {\makecell{SIM ($\uparrow$)}} & {\makecell{WER ($\downarrow$)}} & {\makecell{SIM ($\uparrow$)}} \\
        \midrule
        Seed-TTS~\cite{anastassiou2024seed}     & 1.12 & \textbf{0.796} & 2.25 & \textbf{0.762} \\
        MaskGCT~\cite{wang2024maskgct}      & 2.27 & 0.774 & 2.62 & 0.714 \\
        F5-TTS~\cite{f5tts}       & 1.56 & 0.741 & {\bfseries 1.83} & 0.647 \\
        CosyVoice2~\cite{cosyvoice2}   & 1.45 & 0.748 & 2.57 & 0.652 \\
        Llasa-8B-250k~\cite{ye2025llasa}   & 1.59 & 0.684 & 2.97 & 0.574 \\
        Spark-TTS~\cite{wang2025spark}   & 1.20 & 0.672 & 1.98 & 0.584 \\
        \midrule
        MOSS-TTSD{*}~\cite{moss2025ttsd}   & 3.53 & 0.609 & 9.47 & 0.473 \\
        Vibevoice-1.5B{*}~\cite{peng2025vibevoice}   & 2.65 & 0.689 & 6.62 & 0.570 \\
        FireRedTTS2{*}~\cite{xie2025fireredtts}   & 1.68 & 0.719 & 3.23 & 0.659 \\
        \midrule
        \textbf{SoulX-Podcast}   & \textbf{1.10} & 0.743 & 1.91 & 0.661 \\
        \bottomrule
    \end{tabular}
\end{table*}

\subsection{Podcast Generation}

To evaluate multi-turn, multi-speaker dialogue generation, we compare SoulX-Podcast with representative dialogue TTS systems on the ZipVoice-Dia test set. This benchmark comprises natural multi-turn conversations, enabling assessment of both intelligibility and cross-speaker consistency (cpSIM) in long-form synthesis. As shown in Table~\ref{tab:main_results}, SoulX-Podcast outperforms recent state-of-the-art models on both the Chinese and English subsets. Specifically, it achieves the lowest WER/CER and the highest cpSIM, while maintaining competitive UTMOS scores, demonstrating superior speaker coherence and perceived quality.

\begin{table*}[!hb]
    \centering
    \caption{
        Objective evaluation of multi-speaker TTS systems on ZipVoice-Dia test sets. Arrows indicate the desired direction ($\downarrow$ = lower is better, $\uparrow$ = higher is better). Best values per column are in \textbf{bold}. All model test results below are reproduced from the official release.
    }
    \label{tab:main_results}
    \setlength{\tabcolsep}{1.5mm}
    \begin{tabular}{lcccccc}
        \toprule
        \multirow{2}{*}{\textbf{Model}} & \multicolumn{3}{c}{\textbf{ZipVoice-Dia (zh)}} & \multicolumn{3}{c}{\textbf{ZipVoice-Dia (en)}} \\
        \cmidrule(lr){2-4} \cmidrule(lr){5-7}
        & {\makecell{CER ($\downarrow$)}} & {\makecell{cpSIM ($\uparrow$)}} & {\makecell{UTMOS ($\uparrow$)}} & {\makecell{ WER ($\downarrow$)}} & {\makecell{cpSIM ($\uparrow$)}} & {\makecell{UTMOS ($\uparrow$)}} \\
        \midrule
        ZipVoice-Dia~\cite{zhu2025zipvoice}   & {3.39}  & {0.553}  & \textbf{2.24}   & 3.32 & 0.438 & \textbf{3.10}   \\
        MoonCast~\cite{ju2025mooncast}   & {27.43}  & {0.441}  & {1.76}  & 23.62  & 0.356  & 2.30  \\
        MOSS-TTSD~\cite{moss2025ttsd}  & 8.62  & 0.421  & 1.70   & 8.86   &  0.301   & 2.31      \\
        Vibevoice-1.5B~\cite{peng2025vibevoice}   &{12.87}  &{0.455}  & {1.74}  & {6.58}  & {0.409} & {2.32} \\
        FireRedTTS2~\cite{xie2025fireredtts}   & 3.34   & 0.512  & 1.90   & 5.11   & 0.421 & 2.36  \\
        \midrule
        
        \textbf{SoulX-Podcast} & \bfseries \textbf{2.2}       & \bfseries \textbf{0.599}      & \underline{2.09}       & \bfseries \textbf{2.27}       & \bfseries \textbf{0.484}      & \underline{2.96}     \\
        
        \bottomrule
    \end{tabular}
\end{table*}

\subsection{Evaluation of Paralinguistic Control}

To evaluate the proposed model’s capability for controllable paralinguistic generation, we constructed a dedicated paralinguistic test set. Concretely, we employed GPT-5 to generate 20 test utterances for each of five paralinguistic labels, i.e., \texttt{<|laughter|>}, \texttt{<|sigh|>}, \texttt{<|breathing|>}, \texttt{<|coughing|>}, and \texttt{<throat\_clearing>}. The corresponding audio samples were subsequently synthesized using SoulX-Podcast in monologue speech synthesis mode.

For objective evaluation, we adopted the Qwen-2.5 Omni-FT model~\cite{mai2025mnv17highqualityperformativemandarin} as an automated paralinguistic recognizer. The evaluator was tasked with verifying whether each synthesized utterance contained the target paralinguistic event specified in the prompt. The resulting recognition accuracies are summarized in Table~\ref{tab:recognition_accuracy_en}.

\begin{table}[htbp]
\centering
\caption{Recognition accuracy for different paralinguistic labels.}
\label{tab:recognition_accuracy_en}
\begin{tabular}{l c c c r}
\toprule
Label            & Count  & Correct & Error  & Accuracy \\
\midrule
\texttt{laughter}         & 20     & 20      & 0      & 1.00 \\
\texttt{sigh}             & 20     & 17      & 3      & 0.85  \\
\texttt{breathing}        & 20     & 15      & 5      & 0.75  \\
\texttt{coughing}         & 20     & 14      & 6      & 0.70  \\
\texttt{throat\_clearing} & 20     & 16      & 4      & 0.80  \\
\midrule
Total / Average  & 100    & 82      & 18     & 0.82  \\
\bottomrule
\end{tabular}
\end{table}

As shown in Table \ref{tab:recognition_accuracy_en}, our model achieves a strong overall accuracy of 0.82 in controlling these paralinguistic events. It demonstrates near-perfect control over distinct events like \texttt{<|laughter|>} and high fidelity for \texttt{<|sigh|>} and \texttt{<|throat\_clearing|>}. The primary sources of error appear concentrated in more acoustically subtle or ambiguous events, namely \texttt{<|breathing|>} (0.75) and \texttt{<|coughing|>} (0.70), which may be more challenging for the evaluator model to distinguish.

\subsection{Dialect Generation} 
SoulX-Podcast currently supports three major Chinese dialects: Sichuanese, Henanese, and Cantonese. We evaluate its performance on these dialects in both monologue TTS and dialogue generation settings. The monologue test set includes 1,000 samples per dialect, drawn from internal GPT-generated data as well as SeedTTS, Wenetspeech-Yue-eval~\cite{li2025wenetspeech}, and Wenetspeech-Chuan-eval~\cite{dai2025wenetspeech}. The dialogue test set contains 100 GPT-generated items per dialect.

Dialect-specific ASR systems are used to compute CER, including Wenetspeech-Chuan-ASR~\cite{dai2025wenetspeech} for Sichuanese, TeleSpeech\footnote{\url{https://github.com/Tele-AI/TeleSpeech-ASR}}
 for Henanese, and Wenetspeech-Yue-ASR~\cite{li2025wenetspeech} for Cantonese. As shown in Table~\ref{tab:dialect_evaluation}, SoulX-Podcast achieves consistent speaker similarity across all three dialects, comparable to its performance on Mandarin and English. The relatively high CER values may partly arise from limitations of the ASR systems.

\begin{table*}[!ht]
\centering
\caption{
Performance evaluation of SoulX-Podcast on TTS and dialogue generation across different dialects. Arrows indicate the desired direction ($\downarrow$ = lower is better, $\uparrow$ = higher is better).
}
\label{tab:dialect_evaluation}
\setlength{\tabcolsep}{3mm}
\begin{tabular}{lcccc}
\toprule
\multirow{2}{*}{\textbf{Dialect}} & 
\multicolumn{2}{c}{\textbf{Monologue Test}} & 
\multicolumn{2}{c}{\textbf{Dialogue Test}} \\
\cmidrule(lr){2-3} \cmidrule(lr){4-5}
& CER ($\downarrow$) & SIM ($\uparrow$)
& CER ($\downarrow$) & cpSIM ($\uparrow$) \\
\midrule
Sichuanese & 3.75 & 0.704 & 15.42 & 0.641 \\
Henanese   & 8.14 & 0.705 & 28.06 & 0.647 \\
Cantonese  & 9.77 & 0.680 & 19.50 & 0.627 \\
\bottomrule
\end{tabular}
\end{table*}

\section{Conclusions}



In this work, we introduced SoulX-Podcast, a large language model–driven framework for long-form, multi-speaker, and multi-dialect conversational speech synthesis. Through an interleaved text–speech modeling paradigm, SoulX-Podcast enables the generation of long-form, multi-turn conversational speech with consistent quality and coherence. Experimental results demonstrate that SoulX-Podcast not only excels in multi-turn dialogue synthesis but also generalizes effectively to zero-shot monologue TTS. Its capability to handle multiple Chinese dialects and paralinguistic cues further highlights its versatility and potential as a unified framework for speech generation.

\section{Ethics Statement}
This work focuses on advancing speech synthesis technology through the development of SoulX-Podcast, a large language model–driven framework for multi-speaker and multi-dialect conversational speech generation. All datasets used in this study were either publicly available or synthetically generated, and no personally identifiable information or private recordings were included. 

We acknowledge the potential risks associated with misuse of speech synthesis technology, such as voice spoofing, impersonation, or misinformation. To mitigate these risks, SoulX-Podcast is intended solely for research and responsible development of speech interfaces, and any downstream applications should incorporate appropriate speaker consent, watermarking, and misuse detection mechanisms. We advocate for the transparent, ethical, and human-centric use of speech generation technologies.

\printbibliography
\end{document}